\documentclass[aps,prl,twocolumn,groupedaddress,showpacs]{revtex4}

\usepackage{graphicx}
\usepackage{amsmath}

\begin{document}

\title{Fluctuations and $T_c$ reduction in cuprate superconductors}

\author{J. L. Tallon$^1$, J. G. Storey$^2$ and J. W. Loram$^2$}

\affiliation{$^1$MacDiarmid Institute, Industrial Research Ltd.,
P.O. Box 31310, Lower Hutt, New Zealand.}

\affiliation{$^2$Department of Physics, Cambridge University,
Cambridge CB3 0HE, England.}
\date{\today}

\begin{abstract}
We analyse fluctuations about $T_c$ in the specific
heat of (Y,Ca)Ba$_2$Cu$_3$O$_{7-\delta}$, YBa$_2$Cu$_4$O$_8$ and Bi$_2$Sr$_2$CaCu$_2$O$_{8+\delta}$. The mean-field transition temperature, $T_c^{mf}$, in the absence of fluctuations lies well above $T_c$
especially at low doping where it reaches as high as 150K. We show that phase and amplitude fluctuations set in simultaneously and $T_c^{mf}$ scales with the gap, $\Delta_0$, such that $2\Delta_0/k_BT_c^{mf}$ is comparable to the BCS weak-coupling value, 4.3, for $d$-wave superconductivity. We also show that $T_c^{mf}$ is unrelated to the pseudogap temperature, $T^*$.
\end{abstract}


\maketitle

Many authors have suggested that pairing in the cuprates
begins well above $T_c$. Emery and Kivelsen argued that the low superfluid
density, $n_s$, in the cuprates leads to phase fluctuations below
the mean-field (MF) transition temperature, $T_c^{mf}$, resulting in a
phase-incoherent state with a finite pairing amplitude\cite{Emery}.
In this view phase coherence is not established until a lower
temperature, the observed $T_c$. Experimental support for this picture may be found in the high-frequency optical studies by Corson {\it et al.}\cite{Corson}. Additionally, the underdoped normal state (NS) exhibits a depletion of the density of states (DOS) near the Brillouin zone boundary due to the presence of a
pseudogap\cite{TallonLoram}. The pseudogap seems to close
abruptly at $p_{crit}$=0.19 holes/Cu\cite{TallonLoram}. Some authors
have drawn these two strands together proposing that the
pseudogap corresponds to the phase-incoherent pairing state
between $T_c$ and $T_c^{mf}$\cite{Ong1,Meingast}. The pseudogap $T^*$ line
(below which pseudogap effects are observed) would then correspond
to the doping-dependent $T_c^{mf}$.

Here we analyze the fluctuations in specific heat, $C_P$, to calculate $T_c^{mf}$ and the mean-field jump, $\Delta\gamma^{mf}$, in specific heat coefficient, $\gamma$. We find that at all doping levels $T_c^{mf}$ lies well above the observed $T_c$, reaching as high as 113K for (Y,Ca)Ba$_2$Cu$_3$O$_{7-\delta}$ (Y,Ca-123) and
150K for Bi$_2$Sr$_2$CaCu$_2$O$_{8+\delta}$ (Bi-2212). Our approach is similar to that of Meingast {\it et al.}\cite{Meingast} using thermal expansion data. But, where they identified the pair-fluctuating state with the pseudogap, we show they are distinct. Further, we find that $T_c^{mf}$ and $T^*$ are very different in magnitude and doping dependence, confirming that they are unrelated. Over a wide doping range $T_c^{mf}$ scales with the superconducting gap such that $2\Delta_0/k_BT_c^{mf}$ remains comparable to the weak-coupling BCS $d$-wave value of 4.3.

Even without such an analysis the idea that the pseudogap is a phase incoherent pairing state faces an insurmountable obstacle. If the pseudogap arises merely from thermal phase fluctuations then at $T$=0
there should be no remnant pseudogap effects. But, even at $T$=0 the pseudogap weakens the SC ground state, abruptly reducing the condensation energy\cite{Loram} and superfluid density\cite{Bernhard},
as doping is reduced below $p_{crit}$. The pseudogap thus coexists with SC at
$T$=0 and must be distinct from fluctuation effects above $T_c$.

Firstly, we show that phase and amplitude fluctuations set in simultaneously. Emery and
Kivelsen\cite{Emery} deduce that phase fluctuations become important
above a temperature $T_\theta$ given by $k_BT_\theta \sim A V_0$
where $A \sim 1$ and $V_0$ is the phase stiffness, $V_0 = a\hbar^2n_s(0)/4m^*$.
The length scale, $a$, was defined as $a=\surd\pi\xi$ for
isotropic 3D behavior and $a=\max(d, \surd\pi\xi_\bot)$ for 2D where
$d$ is the mean interlayer spacing. $V_0$ is related to the penetration depth, $\lambda_{ab}$, {\it viz}:
\begin{equation}
\lambda_{ab}^{-2}=\mu_0 e^2(n_s(0)/m^*)= \left(\frac{4\mu_0e^2}{a\hbar^2}\right)V_0 . \label{SF}
\end{equation}
\noindent The condensation energy, $U_0$, is given by
\begin{equation}
U_0= \frac{1}{2} \mu_0 H_c^2=\frac{1}{4} \mu_0 \left(\frac{1}{2\pi\mu_0}\right)^2
\left(\frac{\phi_0}{\lambda_{ab}\xi_{ab}}\right)^2, \label{U0}
\end{equation}
\noindent where the last equality comes from Ginzburg-Landau theory. Combining these, we find
\begin{equation}
k_BT_\theta \sim A V_0 = \left(4A/\pi\right) U_0 \Omega(0) ,
\label{Tphase}
\end{equation}
\noindent where $\Omega(0) = \pi \xi_{ab}^2 a$ is the coherence
volume for Cooper pairs. Following Bulaevskii\cite{Bulaev} we adopt the criterion for amplitude
fluctuations as
\begin{equation}
k_BT_{amp} \sim U_0 \Omega(0) , \label{Tamp}
\end{equation}
\noindent which leads immediately from Eq.~\ref{Tphase} to the
relation
\begin{equation}
k_BT_\theta \sim A V_0 = \left(4A/\pi\right) U_0 \Omega(0) \sim
\left(4A/\pi\right) k_BT_{amp} .
\label{TphTamp}
\end{equation}

As $A\approx0.9$ for 2D\cite{Emery} then the conditions for phase and amplitude
fluctuations are equally restrictive. For a homogeneous system they both must set in simultaneously. We thus question the widely-accepted phase-fluctuation model of Emery and Kivelsen\cite{Emery} and its implementation by Corson {\it et al.}\cite{Corson}. If $T_{\theta}$ and $T_{amp}$ greatly exceed $T_c^{mf}$ then the transition occurs essentially at $T_c^{mf}$. But, if $T_{\theta}$ and $T_{amp}$ are comparable to or less than $T_c^{mf}$ (as is the case) then $T_c$ will be suppressed below $T_c^{mf}$. Between $T_c$ and $T_c^{mf}$ both amplitude and phase will fluctuate, not just the phase. It is our aim to determine how large this $T_c$ suppression is.

The fluctuations in $C_P(T)$ have been analyzed\cite{Schneider,LoramInhomo} by separating $C_P$ into a symmetric fluctuation term, $C_P^{fl}$, and an asymmetric MF term, $C_P^{mf,0}$.
In the 3D-XY model $C_P$ near $T_c$ may be approximated by
\begin{equation}
\Delta C_p=\left\{
\begin{array}{ll}
A^-\ln\mid t\mid + \triangle C_p^{mf,0} & (t\equiv(T/T_c-1)< 0)\\
\\
A^+ \ln\mid t\mid & (t>0).\\
\end{array}\right.
\label{3DXY}
\end{equation}
\noindent $\triangle C_p^{mf,0}$ is the MF step at $T_c$ and $A^-
\approx A^+ = 4k_B/(9\pi^2\Omega(0))$ \cite{Kulic}. While Eq.~\ref{3DXY} is not strictly correct deep in the critical region it captures all the main physical features of the more complex crossover from critical to MF behavior\cite{Salamon} and e.g. accurately represents the critical behavior of He$^4$ at the superfluid transition\cite{Schneider}. A plot of $\triangle C_p$
versus $\ln\mid$$t\mid$ gives two parallel lines offset by
$\triangle C_p^{mf,0}$. In practice this plot exhibits negative
curvature for sufficiently small $\mid$$t\mid$ due to minor transition
broadening. The effect of the resulting spread in $T_c$ was modeled by replacing
$t$ by $t^* = (t^2 + \triangle t^2)^{1/2}$ in the above expressions
for $\triangle C_p$ \cite{Inderhees}.

For Bi-2212 $\Delta C_P^{mf,0}$ was found to collapse rapidly with the opening of the
pseudogap at $p_{crit}$, falling to zero near optimal doping
$p=0.16$ holes/Cu\cite{LoramInhomo}. Below this, $C_P(T)$ is dominated by fluctuations alone and is symmetrical about $T_c$. This is puzzling because the specific heat jump should
remain finite, consistent with the second-order phase transition. We resolve this below.

\begin{figure}
\centerline{\includegraphics*[width=85mm]{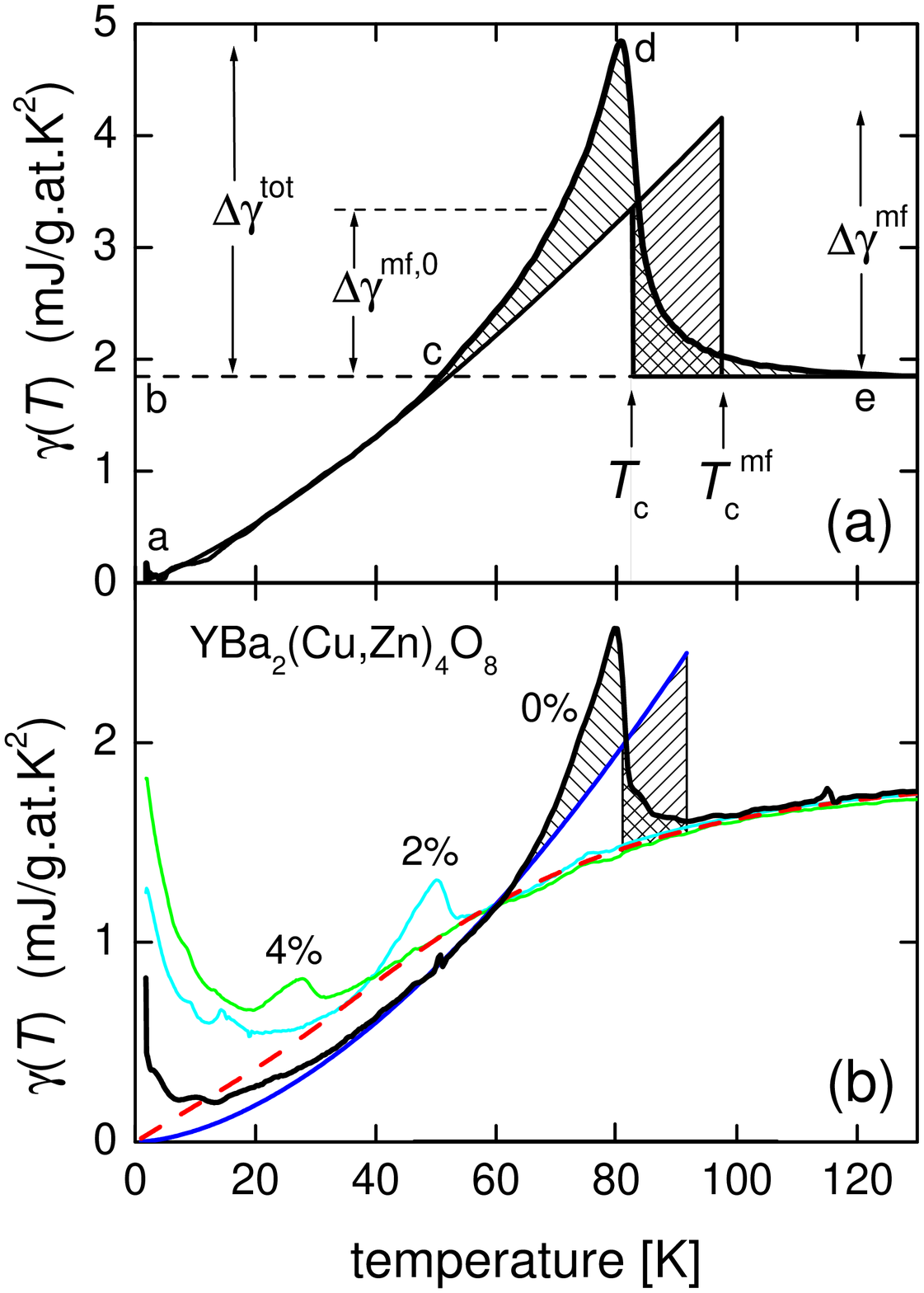}} \caption{\small
(Color online) (a) Analysis of $\gamma(T)$ for Y$_{0.8}$Ca$_{0.2}$Ba$_2$Cu$_3$O$_{6.75}$ showing the MF
component, $\gamma_s^{mf}$, and the fluctuation component above and
below $T_c$. Deduced values are $T_c$=82.79K, $T_c^{mf}$=97.50K, $\Delta
\gamma^{tot}$=3.00, $\Delta \gamma^{mf}$=1.51 and $\Delta \gamma^{mf,0}$=2.31 mJ/g.at.K$^2$. (b) a similar analysis for YBa$_2$Cu$_4$O$_8$ showing $\gamma_s^{mf}$ (blue curve) and the pseudogapped $\gamma_n(T)$ (red dashed curve). The two additional $\gamma(T)$ curves for 2\% Zn (cyan) and 4\% Zn (green) coincide with $\gamma_n(T)$ and thus confirm our pseudogap model. The upturns at low $T$ are due to impurities. Deduced values are $T_c$=81.00K, $T_c^{mf}$=91.92K, $\Delta \gamma^{tot}$=1.11, $\Delta \gamma^{mf}$=0.90 and $\Delta \gamma^{mf,0}$=0.52 mJ/g.at.K$^2$.}
\label{YCagamma}
\end{figure}

Fig.~\ref{YCagamma}(a) shows $\gamma(T)$ reported by Loram {\it et al.}\cite{Loram} for Y$_{0.8}$Ca$_{0.2}$Ba$_2$Cu$_3$O$_{7-\delta}$ with $\delta$=0.25 and $p$=0.186.  The dashed line is $\gamma_n(T)$. Because entropy $S=\int\gamma dT$ there are two entropy balance conditions: (i) the area $abc$ equals the area $cde$. This helps to establish the $T$-dependence of $\gamma_n$ below $T_c$. In this case there is no pseudogap and $\gamma_n(T)$ is constant across the entire $T$-range. For lower doping where the pseudogap is present we use a triangular gap which fills with increasing temperature\cite{TallonLoram}:
\begin{equation}
\gamma_n(T) = \gamma_n(\infty) \left[1 - \vartheta^{-1}\tanh(\vartheta)\ln\left(\cosh(\vartheta)\right)\right] ,
\label{PG}
\end{equation}
\noindent where $\vartheta=E_g/2k_BT$. The second entropy balance condition concerns the fluctuation term which reduces $T_c$ below $T_c^{mf}$. Thus, (ii) the entropy equal to the forward cross-hatched area between $T_c$ and $T_c^{mf}$ equals the fluctuation entropy given by the backward cross-hatched area under the fluctuation term, $\gamma^{fl}$, which includes both critical and Gaussian fluctuations. That is
\begin{equation}
S^{fl} = \int_0^\infty \gamma^{fl} dT = \int_{T_c}^{T_c^{mf}} \left(\gamma_s^{mf} - \gamma_n\right) dT , \label{Scons}
\end{equation}
\noindent This construction enables $T_c^{mf}$ to be estimated. Furthermore, the apparent mean-field step, $\Delta\gamma^{mf,0}$, at $T_c$ is also smaller than the (hypothetical) mean-field step, $\Delta\gamma^{mf}$, that would occur at $T_c^{mf}$. These jumps are defined in the figure. (The same superscript notation is used for the jumps in specific heat $\Delta C_P^{tot}$, $\Delta C_P^{mf,0}$ and $\Delta C_P^{mf}$).

\begin{figure*}
\centering
\includegraphics[width=58mm]{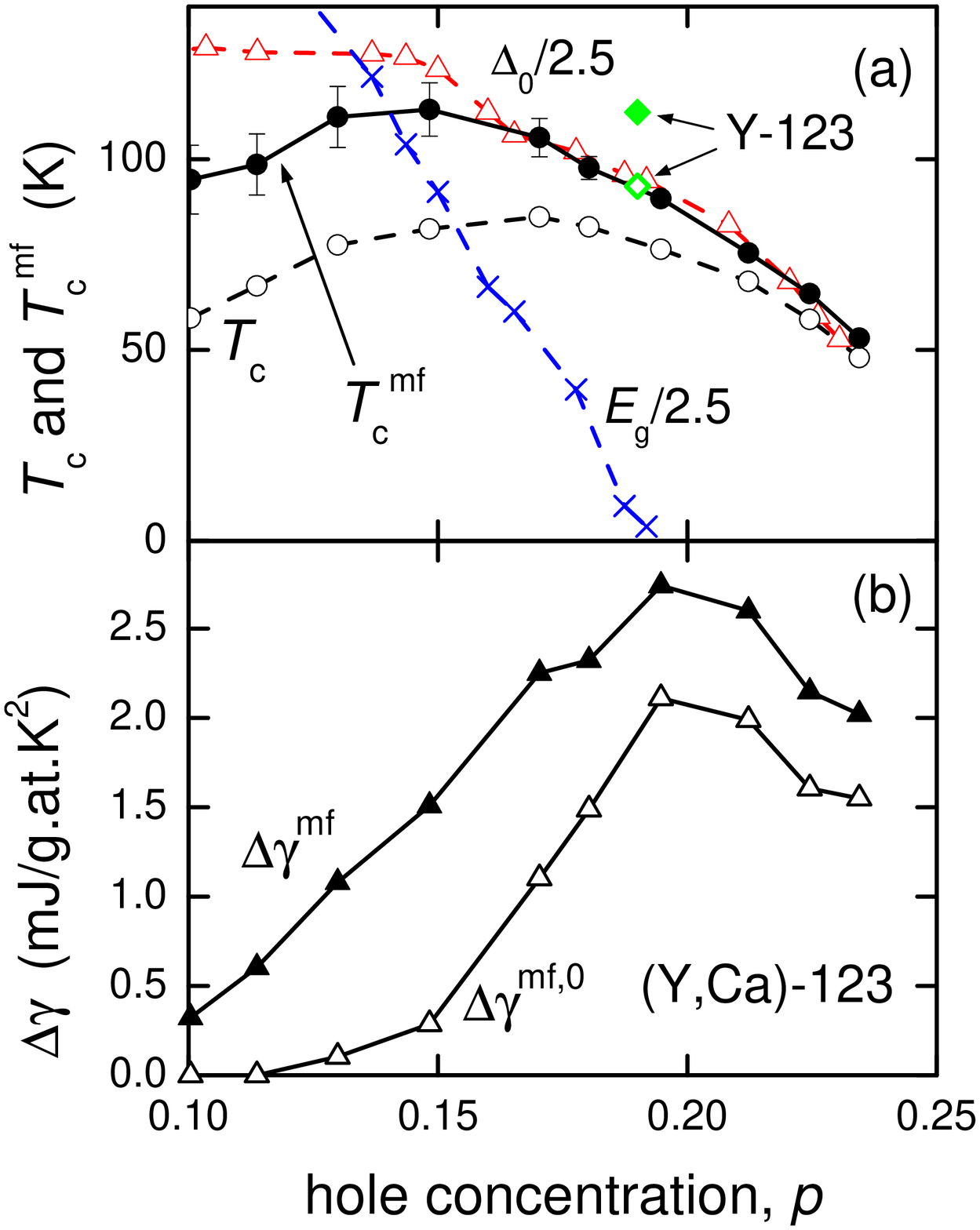}%
\hspace{1mm}%
\includegraphics[width=58mm]{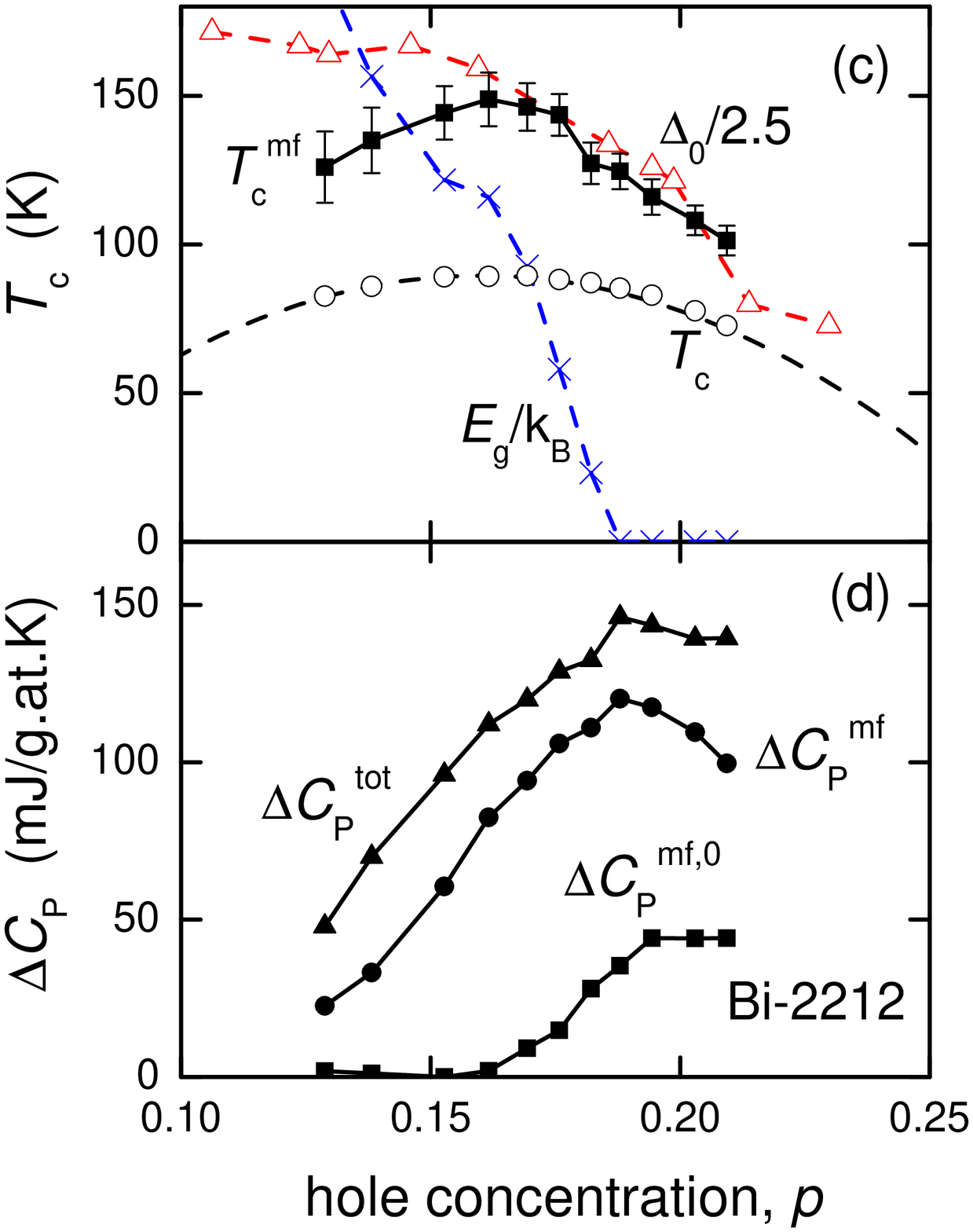}
\hspace{1mm}%
\includegraphics[width=58mm]{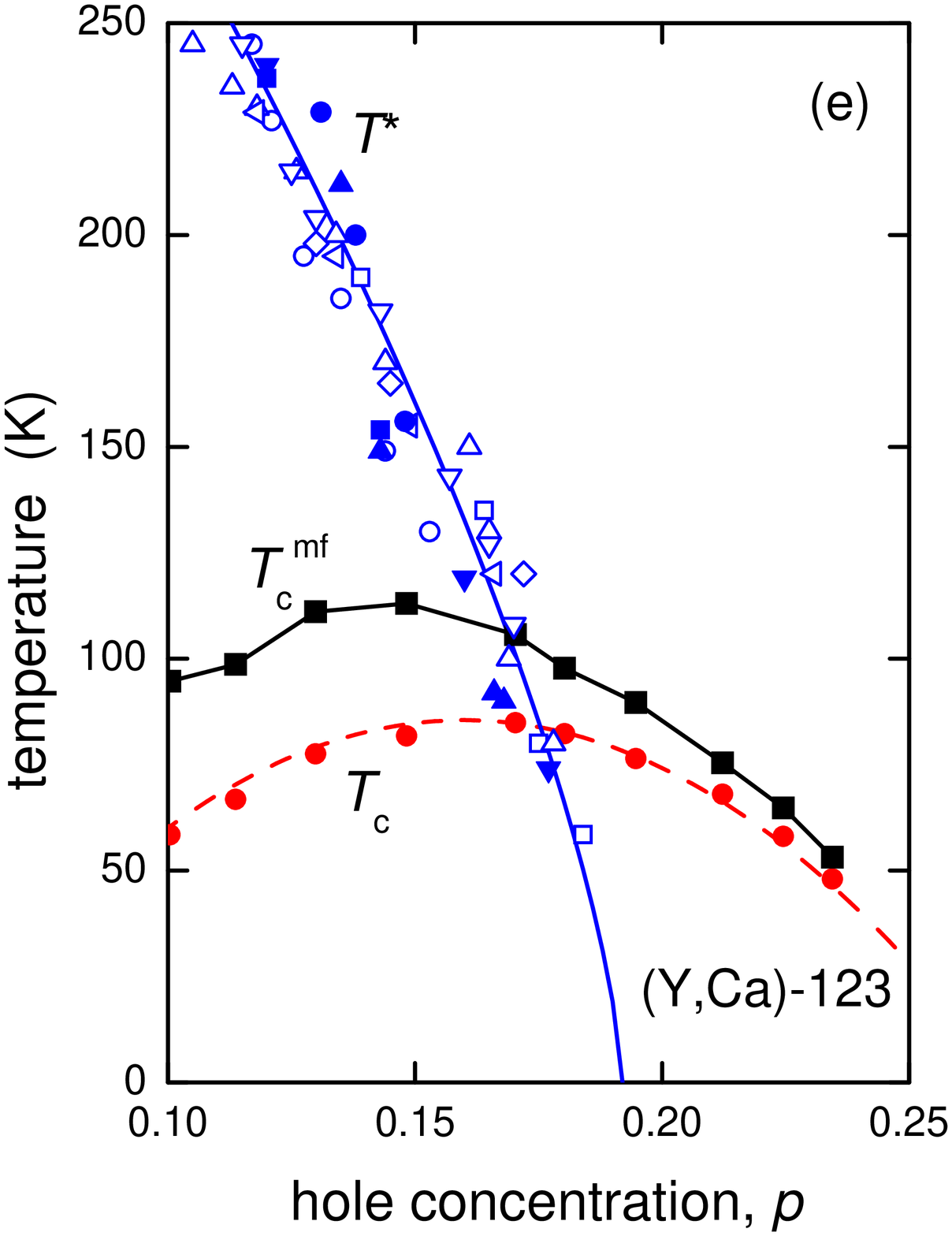}
\caption{\small
(Color online) The doping dependence of evaluated parameters. (a) and (b) show $T_c$, $T_c^{mf}$, $\Delta\gamma^{mf}$ and $\Delta\gamma^{mf,0}$ for Y$_{0.8}$Ca$_{0.2}$Ba$_2$Cu$_3$O$_{7-\delta}$; (c) and (d) show the same for Bi$_2$Sr$_2$CaCu$_2$O$_{8+\delta}$.  In (a) values are also shown for YBa$_2$Cu$_3$O$_{7-\delta}$ (arrows). The SC gap, $\Delta_0$ (red open triangles), and pseudogap $E_g$ (blue crosses) from ref.\cite{Loram} are also plotted, scaled by the factor (1/2.5$k_B$). (c) also shows $\Delta_0$ values for Bi-2212 from $B_{1g}$ Raman (red open triangles). (e) shows $T_c$, $T_c^{mf}$ and $T^*$\cite{Naqib}.} \label{Parameters}
\end{figure*}

We proceed as follows. We combine the first entropy condition with Eq.~\ref{PG} to establish $\gamma_n(T)$. We then plot $C_P$ above and below $T_c$ using Eq.~\ref{3DXY} to determine $\Delta C_P^{mf,0}$ and hence $\Delta\gamma^{mf,0}$\cite{LoramInhomo}. We then construct a power-law fit to $\gamma_s(T)$ at low $T$ that reproduces this value of $\Delta\gamma^{mf,0}$. This is $\gamma^{mf}(T)$ which is slightly superlinear consistent with the predominant $d$-wave gap structure. Lastly, we impose the second entropy condition (Eq.~\ref{Scons}) to deduce $T_c^{mf}$ and $\Delta\gamma^{mf}$. Naturally there are errors inherent in such a construction. But while they grow with underdoping they do not impact on any of our conclusions.

In the example shown in Fig.~\ref{YCagamma} $T_c$ is reduced by ($14.7\pm1.5$) K below $T_c^{mf}$. At the same time $\Delta\gamma^{mf}$=($2.31\pm0.08$) mJ/g.at.K$^2$, significantly more than $\Delta\gamma^{mf,0}$=1.51 mJ/g.at.K$^2$. The analysis was carried out for many samples with different oxygen contents. Values of $T_c$ and $T_c^{mf}$ are plotted in Fig.~\ref{Parameters}(a) along with values of $\Delta \gamma^{mf}$ and $\Delta \gamma^{mf,0}$ in panel (b). A similar analysis was done on specific heat data for Bi-2212\cite{Loram}. Here, instead of using Eq.~\ref{PG} the full bilayer ARPES dispersion was used\cite{StoreyFArc}, thus incorporating the van Hove singularity and pseudogap. The pseudogap was implemented as before\cite{StoreyFArc} using a finite-Fermi-arc model. Results are plotted in Fig.~\ref{Parameters}(c) and (d). Several key conclusions can be made:

(i) Like Meingast {\it et al.}\cite{Meingast} we find $T_c^{mf}(p)$ continues to rise with decreasing doping and only falls at the lowest doping levels. Underdoped samples show a reduction in $T_c$ below $T_c^{mf}$ as large as 35-40K for (Y,Ca)-123 and 60K for Bi-2212, reflecting the larger anisotropy in the latter compound. The shift is also large for pure YBa$_2$Cu$_3$O$_{6.97}$ with $T_c=92.9$K and $T_c^{mf}=112.3$K (see arrowed data points Fig.~\ref{Parameters}(a)).

(ii) while $\Delta\gamma^{mf,0}=0$ at lower doping (and the specific heat anomaly then becomes a pure symmetric fluctuation term) $\Delta\gamma^{mf}$ remains finite in the absence of fluctuations and may only reach zero near the onset of SC at $p\approx0.05$. This removes the puzzle of the seemingly zero MF step.

(iii) Panels (a) and (c) show the pseudogap energy, $E_g$, as previously reported\cite{Loram,StoreyFArc}. Coincident with the abrupt opening of the pseudogap at $p\approx0.19$ there is an abrupt reduction in all values of $\Delta\gamma$ showing that, even after removing fluctuation effects, the pseudogap still plays a decisive role in weakening the condensate and reducing the density of Cooper pairs.

(iv) We compare $T_c^{mf}$ with the $T$=0 SC gap, $\Delta_0$, for the two systems in Fig.~\ref{Parameters}(a) and (c), respectively. For (Y,Ca)-123 values of $\Delta_0$ are from the specific heat\cite{Loram} and for Bi-2212 from the Raman $B_{1g}$ gap\cite{Irwin}. In both cases $2\Delta_0/k_BT_c^{mf}\approx5$ across the entire overdoped region, little more than the $d$-wave mean-field BCS value of 4.3. The old puzzle that $2\Delta_0/k_BT_c$ increases steadily with decreasing doping\cite{Irwin} is now resolved by referencing to $T_c^{mf}$ rather than $T_c$. Eventually, at low doping $T_c^{mf}(p)$ falls below $\Delta_0/2.5k_B$ due to the pseudogap progressively removing spectral weight. If it were not for the pseudogap $T_c^{mf}(p)$ would probably track $\Delta_0/2.5k_B$ across the entire SC domain. The fact that $\Delta\gamma$ is immediately reduced at $p=0.19$ while $T_c^{mf}$ is only gradually reduced by the pseudogap is precisely what is expected in a competing gap scenario\cite{LoramPG}. Thus $T_c$ is reduced both by fluctuations and eventually by the pseudogap once $E_g$ becomes comparable to $\Delta_0$. A corollary is that $T_{c,max}$, the maximum in the $T_c(p)$ phase curve, has no physical significance. It is the $T$=0 $d$-wave SC gap, $\Delta_0$, which is the truly fundamental quantity and in a BCS scenario $T_c^{mf}$ will scale with $\Delta_0$ until the pseudogap opens, as observed.

(v) If we use $\Delta_0$ values for (Y,Ca)-123 from infra-red $c$-axis conductivity\cite{Yu} we obtain $2\Delta_0/k_BT_c^{mf}\approx4.2-4.4$ in even better agreement with the BCS weak-coupling value. In fact, it is probably fair to state that gap magnitudes are not known sufficiently accurately to discount precise agreement with the weak-coupling value.

Fig.~\ref{YCagamma}(b) shows a similar fluctuation analysis on unpublished $\gamma(T)$ data for YBa$_2$(Cu,Zn)$_4$O$_8$ with 0, 2 and 4\% Zn on the planar Cu sites. This shows the rapid suppression of both $T_c$ and the jump in specific heat due to impurity scattering. The high rate of suppression $dT_c/dx= 13$ K/\%Zn is typical of underdoped cuprates and reflects the presence of the pseudogap which enhances the pairbreaking scattering rate due to the reduced DOS\cite{TallonZn}. We use Eq.~\ref{PG} to fit the pseudogapped $\gamma_n(T)$ (shown by the dashed red curve) and the fit is confirmed by the 2\% and 4\% curves for $\gamma(T)$ in Fig.~\ref{YCagamma}(b) for which the NS values coincide with the dashed red curve. The upturns in $\gamma(T)$ at low $T$ are due to a small fraction of impurity and need not concern us.

Next, the values of $\gamma_s^{mf}(T)$ (blue curve) are determined by fitting a power law to $\gamma_s(T)$. The complication of the upturn in the experimental data at low $T$ is averted by insisting on entropy balance such that the area between the dashed red curve and blue curve below the crossing temperature, $T_{cross}=61.2$K, equals the area between the black and dashed red curves above $T_{cross}$. We thus obtain $T_c^{mf}=91.92$K from $T_c=81.00$K; while $\Delta \gamma^{tot}$=1.11, $\Delta \gamma^{mf}$=0.90 and $\Delta \gamma^{mf,0}$=0.52 mJ/g.at.K$^2$. The depression in $T_c$ due to fluctuations is $\Delta T_c=10.92$K, rather less than the value $\Delta T_c=33.5$K obtained for Y$_{0.8}$Ca$_{0.2}$Ba$_2$Cu$_3$O$_{7-\delta}$ at the same doping state. This shortfall is probably due to the enhanced superfluid density in YBa$_2$Cu$_4$O$_8$\cite{Bernhard} which, according to Eqs.~\ref{SF} and ~\ref{Tphase}, will suppress fluctuations. This implies that the gap magnitude is less in the latter compound, perhaps due to the proximity effect between Cu$_2$O$_2$ chains and CuO$_2$ planes which is expected to lower the SC gap magnitude.

There is little data available on the gap magnitude in YBa$_2$Cu$_4$O$_8$ but J\'{a}nossy {\it et al.}\cite{Janossy} have carried out precise measurements of the $T$-dependence of the spin susceptibility below $T_c$ using Gd electron spin resonance. They find an excellent MF $d$-wave fit with maximum (antinodal) gap $\Delta_0$=190K, giving $2\Delta_0/k_BT_c$=4.75. However, by referencing to $T_c^{mf}$ we obtain $2\Delta_0/k_BT_c^{mf}$=4.14, now very close to the weak-coupling value.

Finally, Fig.~\ref{Parameters}(e) compares the various relevant temperature scales, $T_c(p)$, $T_c^{mf}(p)$ and the pseudogap line $T^*(p)$ for Y$_{0.8}$Ca$_{0.2}$Ba$_2$Cu$_3$O$_{7-\delta}$. A central conclusion is that, contrary to some authors\cite{Nakano} the pseudogap line $T^*(p)$ does not merge on the overdoped side with $T_c(p)$, still less with the more fundamental quantity $T_c^{mf}(p)$. This question was effectively put to rest several years ago\cite{Naqib} by an extensive study of the resistivity of high-quality thin films of (Y,Ca)Ba$_2$(Cu,Zn)$_3$O$_{7-\delta}$. The combination of Zn substitution and high magnetic fields allowed $T_c$ to be suppressed so as to expose the evolution of $T^*$ below $T_c$. Panel (e) reproduces these values of $T^*(p)$ (blue data points; solid=films; open=sintered). They extend below the unperturbed $T_c$ value, descending towards zero at $p \approx 0.19$. The solid blue curve is a power-law fit consistent with a terminating quantum critical point\cite{Zaanen}. Importantly Zn substitution and
moderate magnetic fields do not modify $T^*$\cite{Naqib} while they do suppress $T_c$. In this way it is straightforward to distinguish between pseudogap effects and SC fluctuation effects in the transport properties. $T^*$ evidently dips well below $T_c^{mf}$. The temperature scales shown in Fig.~\ref{Parameters}(e) are all of comparable magnitude and it is therefore not surprising that they have been confused in the past.

In summary, we have carried out a fluctuation analysis of
specific heat data to determine the mean-field transition temperature, $T_c^{mf}$, and
the mean-field jump in specific heat coefficient, $\Delta\gamma^{mf}$. $T_c^{mf}$ rises
rapidly above $T_c$ with decreasing doping, reaching values of about 110K for YBa$_2$Cu$_3$O$_{7-\delta}$ and Y$_{0.8}$Ca$_{0.2}$Ba$_2$Cu$_3$O$_{7-\delta}$, and as high as 150K for Bi$_2$Sr$_2$CaCu$_2$O$_{8+\delta}$. This shows the fundamental importance of fluctuations in the cuprates. $\Delta\gamma^{mf}$ remains non-zero across the phase diagram, as it must for a second-order phase transition. The long-standing puzzle that 2$\Delta_0/k_BT_c$ grows with reducing doping is resolved by replacing $T_c$ by $T_c^{mf}$. Across much of the SC phase diagram 2$\Delta_0/k_BT_c^{mf}$ remains close to the weak coupling BCS value. $T^*$ is shown to be distinct from $T_c^{mf}$.

\end{document}